\documentclass[12pt,a4paper]{article}
\usepackage{t1enc}
\usepackage[dvips,final]{graphicx} %<--ked chces vkladat obrazky
\usepackage{tabularx} %<--ked chces robit nieco s tabulkami
\usepackage{amsfonts}

\def\du{\unskip\smash{\lower 1.4ex \hbox{\char34}}\kern-.2ex}
\def\hu{\kern-.2ex\hbox{\char92}}

\newcommand{\bdis}{\begin{displaymath}}
\newcommand{\edis}{\end{displaymath}}
\newcommand{\be}{\begin{equation}}
\newcommand{\ee}{\end{equation}}

\newcommand{\mcal}{\mathcal}

\newtheorem{pr*}[thm]{*}

\begin{document}
\baselineskip=6mm
\newpage

\title{Approximation of the energy levels of the particle on a segment by
energy levels of the particle in monomial potential well}
\author{Michal Demetrian\footnote{{\tt{demetrian@fmph.uniba.sk}}} \\
{\it Comenius University} \\
{\it Mlynsk\' a Dolina M, 842 48, Bratislava IV, Slovak Republic}}
\maketitle

\abstract{This simple text considers an application of Bohr-Sommerfeld quantization rule. It might be of interest for
the students of physics.}

\section{Particle on a segment}

\begin{figure}[h]
\centering
\includegraphics[width=7cm,height=5cm]{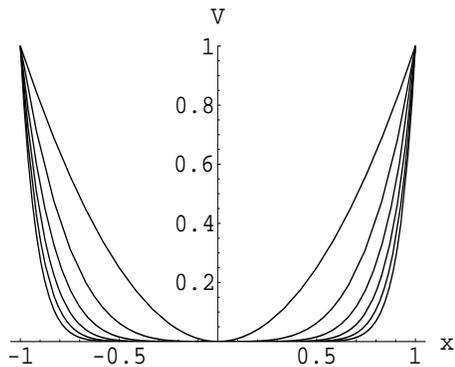}
\caption{There are ploted the first six functions (\ref{potenseq}) with
$A_0^{(k)}=1$ on the
panel.}
\label{potenfig}
\end{figure}

The quantum stationary states and possible values of energy
of the particle on the segment of a line is given by the Schr\" odinger equation
\be \label{sche}
-\frac{\hbar^2}{2m}\frac{{\rm d}^2\psi(x)}{{\rm d}x^2}=E\psi(x)
\ee
and boundary conditions
\be \label{bc}
\psi(0)=\psi(a)=0,
\ee
where $m$ is the mass of the particle and $a$ is the length of the segment. It is well known
that the conditions (\ref{sche}) and (\ref{bc}) lead to the energy eigenfunctions
\be \label{eef}
\psi_n(x)=\sqrt{\frac{2}{a}}\sin\left(\frac{n\pi x}{a}\right),
\quad \mbox{where} \quad n\in\{ 1,2,3,\dots \}
\ee
and energy eigenvalues
\be \label{eev}
E_n=\frac{\pi^2\hbar^2}{2ma^2}n^2\sim n^2.
\ee
Particle on the segment $[0,a]$ can be modeled as particle moving in the line
$(-\infty,\infty)$ in the "intuitive" potential energy
\be \label{poten}
V(x)=\left\{ \begin{array}{l}
0, \ x\in[0,a], \\
\infty, \ x\notin[0,a].
\end{array} \right.
\ee
This kind of potential energy can be, also intuitively as plotted in the
figure (\ref{potenfig}), imagined as "the limit" of the following sequence
of potential energy functions
\be \label{potenseq}
V_k(x)=A_0^{(k)}x^{2k}, \quad k\in\{ 1,2,3,\dots \}
\ee
with some positive constants $A_0^{(k)}$.

Our task is to show that the energy levels (\ref{eev}) can be
approximated by the energy levels of the particle moving in the
potentials (\ref{potenseq}). The tool we are going to use for this
purpose is the well-known Bohr-Sommerfeld quantization rule.

\section{Approximative expression for the energy levels of the particle in the potential well $A_0^{(k)}x^{2k}$}

The Hamiltonian of the particle in the potential well $V=A_0^{(k)}x^{2k}$ is given by
\be \label{ham}
H(p,x)=\frac{p^2}{2m}+A_0^{(k)}x^{2k} .
\ee
Let $\mcal{C}(E)$ be the classical periodic trajectory of the particle with the energy $E$
in the phase space (there are no trajectories of other kind in our case). The
Bohr-Sommerfeld quantization rule chooses from the set of all classical trajectories only
the trajectories for which
\be \label{bsrule}
\oint_{\mcal{C}(E)}p{\rm d}x=2\pi\hbar(n+\gamma_n),
\ee
where $n$ is natural number and $\gamma_n$ is the quantity of the order of unity.
For our purpose, it is sufficient to write down the rule in less accurate form
\be \label{bsrule1}
\oint_{\mcal{C}(E)}p{\rm d}x=2\pi\hbar n.
\ee
In this way, Bohr-Sommerfeld quantization rule gives us the method to determine
approximately the energy levels of the particle by solving the equation (\ref{bsrule1})
with the unknown $E$ and parameter $n$. The accuracy of this formula grows with growing $n$.
In our case, the particle with the energy $E$ moves
in the range $[-x_M(E),x_M(E)]$, where the energy conservation law determines $x_M(E)$ as
\be \label{xm}
x_M(E)=\left(\frac{E}{A_0^{(k)}}\right)^{\frac{1}{2k}} .
\ee
Therefore, the left-hand side of Eq. (\ref{bsrule1}) reads
\begin{eqnarray*}
& &
\oint_{\mcal{C}(E)}p{\rm d}x=4\int_0^{x_M}p(x){\rm d}x=4\sqrt{2m}\int_0^{x_M}
\sqrt{E-A_0^{(k)}x^{2k}}{\rm d}x= \\
& &
4x_M\sqrt{2mE}\int_0^1\sqrt{1-y^{2k}}{\rm d}y=
\frac{4x_M\sqrt{2mE}}{2k}\int_0^1(1-t)^{1/2}t^{\frac{1}{2k}-1}{\rm d}t= \\
& &
\frac{2}{k}(2m)^{1/2}\left(A_0^{(k)}\right)^{-\frac{1}{2k}}E^{\frac{1}{2}+\frac{1}{2k}}
\mcal{B}\left(\frac{3}{2},\frac{1}{2k}\right),
\end{eqnarray*}
where $\mcal{B}$ is the Euler's beta-function. Above written
equalities together with (\ref{bsrule1}) allow for expressing
energetic levels of the system in closed form
\be \label{spect}
E_n^{(k)} =\left[\frac{\pi\hbar
k}{(2m)^{1/2}}\right]^{\frac{2k}{k+1}}
\left(A_0^{(k)}\right)^{\frac{1}{k+1}}
\left[\frac{1}{\mcal{B}\left(\frac{3}{2},\frac{1}{2k}\right)}\right]^{\frac{2k}{k+1}}
n^{\frac{2k}{k+1}} .
\ee
We are interested in asymptotic behavior of
$E_n^{(k)}$ at fixed $n$ and $k$ running to infinity. By making
use the facts
\bdis
\mcal{B}(x,y)=\frac{\Gamma(x)\Gamma(y)}{\Gamma(x+y)} \quad
\mbox{and} \quad \Gamma(x)=\frac{1}{x}-\gamma_E+\mcal{O}(x) \ \mbox{for}\ x\to 0^+ ,
\edis
where $\gamma_E$ is the Euler's constant, and expressing the
constant $A_0^{(k)}$ in the form appropriate to keep the width of
the well equal to a constant $2a$
\bdis
A_0^{(k)}=\frac{V_0}{a^{2k}}, \quad V_0=const, \quad V_0>0,
\edis
we can derive that for $k\to\infty$
\bdis
E^{(k)}_n=\left[\frac{\pi^2\hbar^2}{2ma^2}k^2+k+\mcal{O}(1)\right]
\left[\Gamma\left(\frac{1}{2k}\right)\right]^{-\frac{2k}{k+1}}\left[
n^2+\mcal{O}\left(\frac{1}{k}\right)\right]
\edis and therefore
\be \label{limen}
\lim_{k\to\infty}E_n^{(k)}=\frac{\pi^2\hbar^2}{2m\left(2a\right)^2}n^2
\ee
as it should be.

\end{document}